\documentclass[preprintnumbers,showpacs,amsmath,amssymb,floatfix,prd,twocolumn,superscriptaddress,nofootinbib]{revtex4}
\usepackage{bbm}
\usepackage{graphicx}
\usepackage{epsfig}
\usepackage{bm}
\usepackage{amsfonts}
\usepackage{epstopdf}

\begin{document}

\title{Thermodynamic  Origin of the Cardassian Universe}

\author{Chao-Jun Feng}
\email{fengcj@shnu.edu.cn} \affiliation{Shanghai United Center for Astrophysics (SUCA), \\ Shanghai Normal University,
    100 Guilin Road, Shanghai 200234, P.R.China}

\author{Xin-Zhou Li}
\email{kychz@shnu.edu.cn} \affiliation{Shanghai United Center for Astrophysics (SUCA),  \\ Shanghai Normal University,
    100 Guilin Road, Shanghai 200234, P.R.China}

\author{Xian-Yong Shen}
\email{1000304237@smail.shnu.edu.cn } \affiliation{Shanghai United Center for Astrophysics (SUCA), \\ Shanghai Normal University,
    100 Guilin Road, Shanghai 200234, P.R.China}

\begin{abstract}
In the Cadassian universe,  one can explain the acceleration of the universe without introducing dark energy component. However,   the dynamical equations of this model can not be directly obtained from the action principle. Recently, works on the relation between thermodynamics and gravity indicates that gravity force may not be the fundamental force.  In this paper, we study the thermodynamics of the Cardassian universe, and regard it as the origin of this cosmological model. We find that the corresponding entropy obeys ordinary area law when the area of the trapping horizon is small, while it becomes a constant when area is going to be large in the original and modified polytropic Cardassian model, and it has a maximum value in the exponential one. It seems that the Cardassian universe only contains finite information according to the holographic principle, which states that all the information in the bulk should be encoded in the boundary of the bulk.   
\end{abstract}


\maketitle


\section{Introduction}\label{sec:intro}

Recently,  the relation between the laws of thermodynamics and that of gravity has aroused great interest. In Hawking's  work \cite{Hawking:1974sw}, it shows that black holes have thermal radiation with a temperature determined by its surface gravity due to the quantum effect, which implies that these two branches of physics  may have underly relations, namely, they are unified. The authors in ref.~\cite{Hayward:1997jp} proposed a unified first law of black hole dynamics and relativistic thermodynamics in spherically symmetric general relativity, see also \cite{Hayward:1993wb, Hayward:1994bu, Hayward:1998ee}.  By using  the unified law in the framework of Friedmann-Robertson-Walker (FRW) universe, the authors in \cite{Cai:2006rs} find the relation between the Friedmann equation and the first law of thermodynamics on the ``inner'' trapping horizon. They also find that the Clausius relation holds for the Gauss-Bonnet and Lovelock theory by treating the higher derivative terms as an effective energy-momentum tensor, but the Clausius relation never holds for the scalar-tensor theory, which implies that this is a system with non-equilibrium thermodynamics.  These results indicate that the gravity may be not one of the  fundamental forces and it is only another view of thermodynamics. In this spirit, we can regard  the thermodynamics law as  the fundamental law and   it will give the corresponding gravity theory. Thus, one can study the thermodynamics of some   cosmological models to  pursue their origins. For related discussions, see also \cite{Akbar:2006mq, Akbar:2006er, Cai:2008ys, Cai:2008mh, Zhang:2010hi, Hu:2010tx, Cao:2010vj}.

Up to now, there are many  kinds of dark energy models and modified gravity theories proposed to explain the current accelerating
expansion of the universe, which has been confirmed by the observations like Type Ia supernovae (SNe Ia), CMB and SDSS
et al. The dark energy models assume the existence of an energy component with negative pressure in the universe, and
it dominates and accelerates the universe at late times. The cosmological constant seems the best candidate of dark
energy, but it suffers the fine tuning problem and coincidence problem, and it may even have the age problem
\cite{Feng:2009jr}. To alleviate these problems, many dynamic dark energy models were proposed. However, people still
do not know what is dark energy.

Since the Einstein general gravity theory has not been checked in a very large scale, then one does not know whether
this gravity theory  is suitable or not for studying the observational data like SNe Ia, and maybe the accelerating
expansion of universe is due to the gravity theory that differs from the general gravity. Thus, many modified gravity
theories like $f(R)$, DGP et al. are proposed to explain the accelerating phenomenology. The Cardassian model is a kind
of model in which the Friedmann equation is modified by the introduction of an additional nonlinear term of energy
density and in this model one does not need dark energy component also. However, one can not directly find  its origin from the first principle. 

As we mentioned before, the gravity may be not  a fundamental theory, and one can at least study the thermodynamics of a cosmological model to pursue its origin. So, in this paper, we will  study the thermodynamics of the Cardassian universe, and regarded it as its origin. We find that the corresponding  entropy obeys ordinary area law $S=A/(4G)$ on the trapping horizon when $A$ is small, where $A$ is the  the area of the trapping horizon, and it becomes a constant when $A$ is going to be large in the original (OC) \cite{Freese:2002sq} and modified polytropic Cardassian model  (MPC) \cite{Wang:2003cs}, while it has a maximum value in the exponential model (EC) \cite{Liu:2005cm, Sun:2009pb}. It seems that the Cardassian universe only contains finite information according to the holographic principle, which states that all the information in the bulk should be encoded in the boundary of the bulk.

This paper is organized as follows: In Section \ref{sec:firstlaw}, we will give a briefly review on the unified first law and its application. In Section \ref{sec:car}, we will study the thermodynamics of the Cardassian universe including the OC, MPC and EC models, for recent works on the Cardassian universe, see \cite{Feng:2009zf, Liang:2010vu, Wang:2009xq, Wang:2009ra}. In the last section, we will give some discussions and conclusions.

%
%

\section{Briefly review on the unified first law}\label{sec:firstlaw}
Einstein equations can be written in a form called "unified first law" based on the general definition of black hole dynamics on trapping horizon, which was proposed by Hayward \cite{Hayward:1997jp, Hayward:1993wb, Hayward:1994bu, Hayward:1998ee} and developed by Cai et al \cite{Cai:2006rs, Akbar:2006mq, Akbar:2006er, Cai:2008ys, Cai:2008mh}. In the rest of this section, we will give a brief review on this unified first law in the $3+1$-dimensional spherical symmetric spacetime, in which the metric could be locally written in the double-null form as
\begin{equation}\label{double_null}
	ds^{2}=-2e^{-f}d\xi^{+}d\xi^{-}+r^{2}d\Omega^{2} \,,
\end{equation}    
where $d\Omega^{2}$ is the line element of the $2$-sphere with unit radius, $r$ and $f$ are functions of ($\xi^{+},\xi^{-}$).  So, each symmetric sphere has two preferred normal directions, namely the null directions $\partial /\partial\xi^{\pm}$, which will be assumed future-pointing in the following.  And also, we will assume the spacetime is time-orientable.  The expansions of the radial null geodesic congruence are defined by 
\begin{equation}\label{double_null}
            \theta_{\pm}=2r^{-1}\partial_{\pm}r \,,
\end{equation}
where $\partial_{\pm}$ denotes the coordinates derivative along $\xi^{\pm}$. The expansion measures whether the light rays normal to the sphere are diverging $(\theta_{\pm} > 0)$ or converging $(\theta_{\pm} < 0)$, namely, whether the sphere is increasing or decreasing in the null directions. Note that, although the value of $\theta_{\pm}$ will change with geometries, its sign will not, and the only invariants of the metric and its first derivative are functions of $r$ and $e^f \theta_{+}\theta_{-}$, or equivalently $g^{ab}\partial_{a}r\partial_{b}r = -\frac{1}{2}e^{f}\theta_{+}\theta_{-}$, which has an important physical and geometrical meaning: a sphere is said to be trapped (untrapped) if $\theta_{+}\theta_{-} >0  ( \theta_{+}\theta_{-} <0 ) $,  and if $\theta_{+}\theta_{-}=0$, it is a marginal sphere. 

Considering non-stationary black holes, Hayward has proposed that the future outer trapping horizons defined as the the closure of a hypersurface foliated by future or past, outer or innner marginal sphere is taken as the definition of black holes, since the horizon possess various properties which are often intuitively ascribed to black hole including confinement of observers and analogues of the zeroth, first and second law of thermodynamics. However, in the case of FRW universe,  one should take the  future inner trapping horizon defined by
\begin{equation}
	\theta_{+} = 0\,, \quad \theta_{-} < 0\,, \quad \partial_{-}\theta_{+} >0 \,,
\end{equation}
 as a system on which the thermodynamics will be established, since the surface gravity is negative on the cosmological horizon.

In a spherical symmetric spacetime, one can obtain the total energy inside the sphere with radius $r$ by calculating the Minsner-Sharp energy given by 
\begin{equation}
	E = \frac{r}{2G}\bigg(1-g^{ab}\partial_{a}r\partial_{b}r \bigg) =  \frac{r}{G}\bigg(\frac{1}{2}-g^{+-}\partial_{+}r\partial_{-}r \bigg) \,,
\end{equation}
 which is a pure geometric quantity and has much better properties than the other definitions of energy  when one consider the case of non-stationary spacetime. The relation between the  Minsner-Sharp energy and others could be found in ref.~\cite{Hayward:1994bu} .   There are also two invariants constructed from the  energy-momentum tensor $T^{\mu\nu}$: 
\begin{equation}
	W =  - \frac{1}{2} g_{ab}T^{ab} \,   = -g_{+-}T^{+-} \,, 
\end{equation}
which are called the work density, and $\Psi$ called  the energy flux vector (also called the energy-supply  vector), whose components are
\begin{equation}
	\Psi_{a} = T^{b}_{a} \partial_{b} r + W \partial_{a} r \,.
\end{equation}
Here and in the following,  $a, b$ denotes the two dimension space normal to the sphere. With the help of the definition of Minsner-Sharp energy and the above two  quantities,  one can write the $(0,0)$ component of Einstein equations as a ``unified first law'' :
\begin{equation}\label{ufl}
	dE = A\Psi + WdV \,,
\end{equation}
where $A = 4\pi r^{2}$ and $V = \frac{4\pi}{3} r^{3}$.  This unified law contains rich information, e.g. by projecting the unified first law along the trapping horizon, we can obtains the first law of black hole thermodynamics,which has the form \cite{Cai:2006rs}
\begin{equation}\label{uflofbh}
            \langle dE,z \rangle =  \frac{\kappa}{8\pi G}\langle dA,z \rangle + \langle  WdV,z \rangle   \,,
\end{equation}    
where $\kappa$ defined by 
\begin{equation}\label{surg}
	\kappa = \frac{1}{2} \nabla^{a} \nabla_{a} r \,,
\end{equation}
 is the surface gravity of the trapping horizion. Here $z=z^{+}\partial_{+}+z^{-}\partial_{-}$ is a vector tangent to the trapping horizon, and it should be noticed that  by definition of the horizon $\partial_{+}r = 0$, one has 
\begin{equation}\label{zpmpre}
	z^{a}\partial_{a}(\partial_{+}r) = z^{+}\partial_{+}\partial_{+} r + z^{-} \partial_{-}\partial_{+}r = 0 
\end{equation}
on the trapping horizon, then 
\begin{equation}\label{zpm}
	\frac{z^{-}}{z_{+}}  = - \frac{\partial_{+}\partial_{+} r}{\partial_{-}\partial_{+} r} \,.
\end{equation}
Also note that by taking the Einstein equations  $\partial_{+}\partial_{+} r = -4\pi r T_{++}$, see ref.~\cite{Hayward:1994bu} and the definition of the surface gravity in Eq.~(\ref{surg}), one can easily finds 
\begin{equation}
	 \langle A\Psi,z \rangle = \frac{\kappa}{8\pi G}\langle dA,z \rangle \,,
\end{equation}
which is the Clausius relation in the version of black hole thermodynamics, see the first term on the right side of  Eq.~(\ref{uflofbh}). The left side of the above equation is nothing but the heat flow $\delta Q$, and the right side has the form $TdS$, if one identities the temperature $T=\kappa/2\pi$ and the  entropy $S = A/4G$. So, in Einstein theory, the ``unified first law'' also implies the Clausius relation, and this relation is also hold in the Gauss-Bonet and Lovelock gravity theories  by treating the higher derivative terms as an effective energy-momentum tensor. But, in the scalar-tensor theory, this relation is no longer hold  due to some non-equilibrium thermodynamical properties.  In the next section, we will study the thermodynamics of the Cardassian universe, and calculate the corresponding entropy.   

\section{Thermodynamics of the Cardassian universe}\label{sec:car}

The spacetime of the Cardassian universe is described by the FRW metric, which could be written in the form of 
\begin{equation}\label{FRW}
            ds^{2}=h_{ab}dx^{a}dx^{b}+\tilde{r}d\Omega^{2} \,,
\end{equation}
 where $x^{0}=t$, $x^{1}=r$ and  $\tilde{r}=a(t)r$, which is the radius of the sphere while $a(t)$ is the scale factor. Defining
 \begin{equation}
	   d\xi^{\pm}=-\frac{1}{\sqrt{2}}\left(dt \mp \frac{a}{\sqrt{1-kr^2}}dr\right) \,,
\end{equation}
where $k$ is the spacial curvature, the metric could be rewritten as a double-null form
\begin{equation}
	ds^2=-2 d\xi^+d\xi^- +\tilde{r}^2d\Omega^2 \,,
\end{equation}
then we get the trapping horizon $\tilde r_{T}$ by solving the equation $\partial_{+} \tilde r |_{\tilde r= \tilde r_{T}} = 0$ as 
\begin{equation}
	\tilde r_{T} = \left(H^{2}+ \frac{k}{a^{2} } \right )^{-1/2} \,,
\end{equation}
which has the same form of the apparent horizon.  Thus, one can get the surface gravity $\kappa = -(1-\epsilon)/\tilde r_{T}$, where  we have defined
\begin{equation}
	\epsilon \equiv \frac{\dot{\tilde r}_{T}}{2H \tilde r_{T}} \,.
\end{equation}
One can also check that $\partial_{-}\tilde r_{T} <0$ indicating the trapping horizon is future.  By using Eq.~(\ref{zpmpre}) and after a direct calculation, one can get $z^{-} = \epsilon/(1-\epsilon)$ when $z^{+} = 1$ is chosen. Then, in the $(t, r)$ coordinates, the project vector is given by $z= \partial_{t} - (1-2\epsilon)Hr \partial_{r}$.
 
In the  Cardassian universe, the Friedmann equation is modified as  
\begin{equation}\label{modFried}
	H^{2}+\frac{k}{a^{2}}=\frac{8\pi G}{3}g(\rho_{m})  = \frac{8\pi G}{3} (\rho_{m} + \rho_{e})\,,
\end{equation}
where $g$ is some function of the energy density of matter and we have defined the effect energy density $\rho_{e} = g(\rho_{m}) - \rho_{m}$. Using the continuity equations, we can obtain the effective pressure corresponds to the effective energy density
\begin{equation}
	p_{e}=(\rho_{m}+p_{m})g'(\rho_{m})-g(\rho_{m})-p_{m} \,.
\end{equation}
where the prime denotes the derivative with respect to $\rho_{m}$.  Then, we get the associated work density $W_{e}$ and  energy-supply vector $\Psi_{e}$ as
\begin{eqnarray}
	W_{e}    &=& \frac{1}{2} \bigg[ 2g-\rho_{m}+p_{m} -(\rho_{m}+p_{m})g' \bigg] \,, \\
	\Psi_{e} &=& \frac{1}{2} (g'-1)(\rho_{m}+p_{m})(-H\tilde r_{T} dt + a dr) \,.
\end{eqnarray}
Therefore, we obtain 
\begin{equation}\label{heatf2}
	\delta Q_{e} = \langle A\Psi_{e}, z\rangle =  \frac{\kappa A H\epsilon }{2\pi G}\left( \frac{g'-1}{g'} \right) = T\left( \frac{g'-1}{4G g'} \right) \langle dA, z\rangle\,,
\end{equation}
where we have used the relation
\begin{equation}
	\dot H - \frac{k}{a^{2}} = -\frac{2\epsilon}{\tilde r_{T}^{2}} = -4\pi G g' (\rho_{m} + p_{m}) \,.
\end{equation}
Here, we have also identified  $T = \kappa/2\pi$.  And, for the heat flow of pure matter, we also have
\begin{equation}\label{heatf}
	\delta Q_{m} = \frac{\kappa}{8\pi G}\langle dA, z\rangle - \langle A\Psi_{e}, z\rangle =  \frac{T}{4Gg'} \langle dA, z\rangle\,,
\end{equation}
So, when $g_{m} = \rho_{m}$, the above equation reduces to the Clausius relation in the unmodified Friedmann model.  In the following, we will focus on some concrete Cardassian models, namely, the original Cardassian model (OC), the modified polytropic Cardassian model (MPC) and the exponential model (EC). In these models, the function $g(\rho_{m})$ takes different forms, which will reduce to $\rho_{m}$ in the early universe and  differ from the unmodified Friedmann universe at redshift $z<\mathcal{O}(1)$, during which it will gives rise to accelerated expansion.
 
\subsection{OC model}
In this model, the function $g(\rho_{m})$ is given by
\begin{equation}\label{oc model}
	g(\rho_{m}) = \rho_{m} \left[1+ \left(\frac{\rho_{m}}{\rho_{c}}\right)^{n-1}\right] = \rho_{m}\big[1+f_{o}(\rho_{m})\big]\,,
\end{equation}
where $\rho_{c}$ is a character energy density in the Cardassian universe and the parameter $n$ is assumed to satisfy $n<2/3$ to give rise to a acceleration of the universe.  Here we have defined the function $f_{o}(\rho_{m}) = (\rho_{m}/\rho_{c})^{n-1}$. Thus, by using Eq.~(\ref{heatf}), the heat flow of pure matter  is given by 
\begin{equation}
	\delta Q_{m} =  \frac{T}{4G(1+nf_{o})} \langle dA, z\rangle = T\langle dS_{m}, z \rangle\,,
\end{equation}
where the entropy is obtained by 
\begin{equation}\label{entropy oc}
	dS_{m} = \frac{dA}{4G(1+nf_{o})} \,,
\end{equation}
which reduces to the usual relation $dS = dA/ 4G$ in the limit of  $f_{o}\rightarrow 0$. However, when $f_{o}$ is large enough, the entropy becomes a constant that independent of the surface area, which means there is no heat flow of pure matter on the trapping horizon, but this time $dS_{e} = dA/4G$. From  Friedmann equation (\ref{modFried}),  we also have the following constraint equation
\begin{equation}\label{cons1}
	f_{o}^{\frac{1}{n-1}} (1+f_{o}) = \frac{3}{2GA\rho_{c}} \,.
\end{equation}
Therefore, we get the entropy by integrating Eq.~(\ref{entropy oc})
\begin{eqnarray}
\nonumber
	S_{m} &=& - \frac{3}{8G^{2}\rho_{c}(n-1)}\int  f_{o}^{-\frac{n}{n-1}}(1+f_{o})^{-2}  df_{o}  \\
	&=& \frac{A}{4G} \bigg(1+f_{o}\bigg) {}_2F_{1}\bigg[2, \frac{-1}{n-1}, \frac{n-2}{n-1}, -f_{o}\bigg] \,, \label{entropy oc2}
\end{eqnarray}
up to some integration constant.  Here ${}_2F_{1}$ is the hypergeometric function and Eq.~(\ref{cons1}) gives the relation between $f_{o}$ and $A$.
\begin{figure}[h]
\begin{center}
\includegraphics[width=0.4\textwidth]{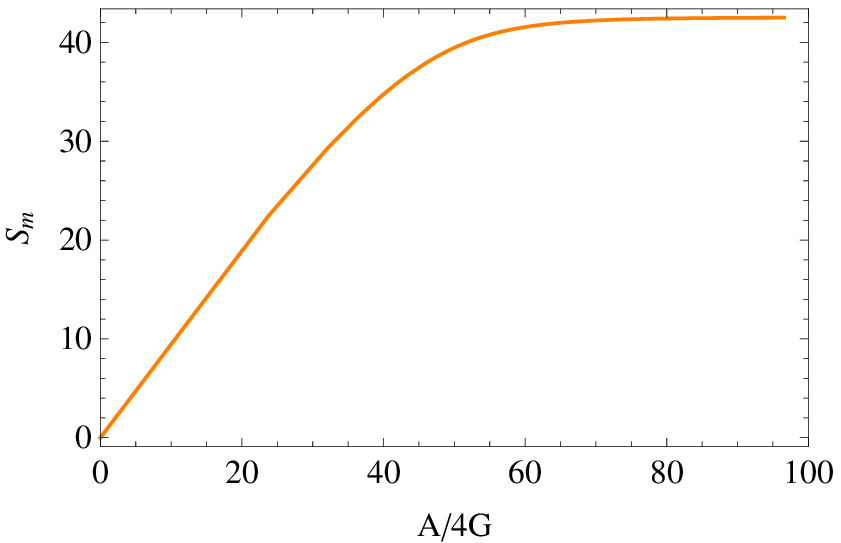} 
\qquad
\includegraphics[width=0.4\textwidth]{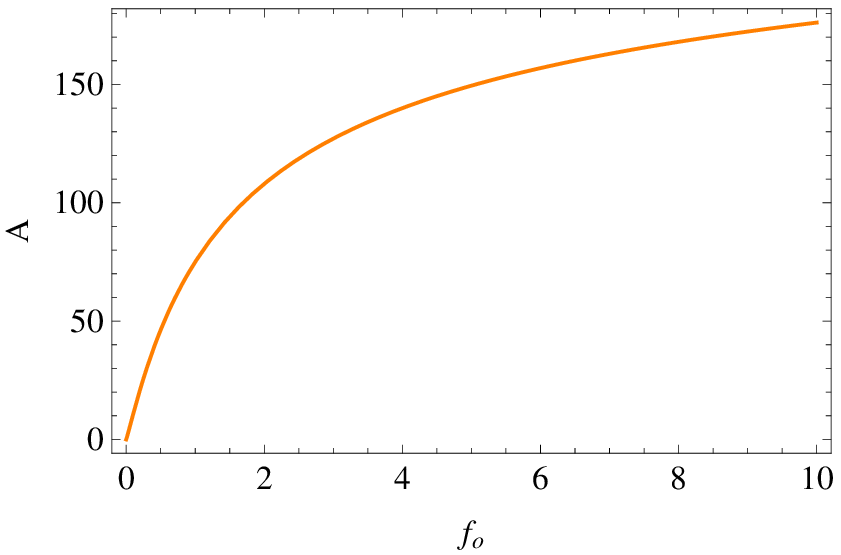} 
\caption{\label{fig::eoc1}Top: The entropy of the original Cardassian universe as the function of the surface area with parameter $n=0.1$. Bottom: The area of the trapping horizon with respect to function $f_{o}$ with the same parameters. }
\end{center}
\end{figure}
To illustrate the relation between  the entropy and the surface area, we plot an example with parameter $n=0.1$ in Fig.~\ref{fig::eoc1}, in which it shows that when $f_{o}$ is small, the entropy satisfies the usual area law $S = A/4G$, while $f_{o}$ is large, it becomes a constant as we see in Eq.~(\ref{entropy oc}). Actually, from Eq.~(\ref{entropy oc2}), one can obtain that 
\begin{equation}\label{oc f}
	S_{m}\big|_{f_{o}\rightarrow \infty} = \frac{3}{8G^{2}\rho_{c}} \Gamma\left(\frac{n-2}{n-1}\right) \Gamma\left(\frac{2n-1}{n-1}\right) \,,
\end{equation}
where $\Gamma$ denotes  Gamma functions.  

\subsection{MPC model}
The modified polytropic Cardassian model can be obtained by introducing an additional parameter $q>0$ into the original Cardassian model, in which the function $g(\rho_{m})$ is given by
\begin{equation}\label{mpc model}
	g(\rho_{m}) = \rho_{m} \left[1+ \left(\frac{\rho_{m}}{\rho_{c}}\right)^{q(n-1)}\right]^{\frac{1}{q}}  =  \rho_{m}\big[1+f_{q}(\rho_{m}) \big]^{\frac{1}{q}} \,,
\end{equation}
where $f_{q} = (\rho_{m}/\rho_{c})^{q(n-1)}$, and when $q=1$, it reduces to the original model. Thus, by using Eq.~(\ref{heatf}), the heat flow of pure matter  is given by 
\begin{equation}
	\delta Q_{m} =  \frac{T}{4G(1+nf_{q})(1+f_{q})^{\frac{1}{q} - 1}} \langle dA, z\rangle = T\langle dS_{m}, z \rangle\,,
\end{equation}
where the entropy is obtained by 
\begin{equation}\label{entropy mpc}
	dS_{m} = \frac{dA}{4G(1+nf_{q})(1+f_{q})^{\frac{1}{q} - 1}} \,,
\end{equation}
which also reduces to the usual relation $dS = dA/ 4G$ in the limit of  $f_{q}\rightarrow 0$, and the entropy becomes a constant that independent of the surface area when $f_{q}$ is large enough. From  Friedmann equation (\ref{modFried}),  we also have the following constraint equation
\begin{equation}\label{cons2}
	f_{q}^{\frac{1}{q(n-1)}} (1+f_{q})^{\frac{1}{q}} = \frac{3}{2GA\rho_{c}} \,.
\end{equation}
Therefore, we get the entropy by integrating Eq.~(\ref{entropy mpc})
\begin{eqnarray}
\nonumber
	S_{m}  &=& - \frac{3}{8G^{2}\rho_{c}q(n-1)}\int  f_{q}^{-\frac{1}{q(n-1)}-1}(1+f_{q})^{-\frac{2}{q}}  df_{q}   \\
\nonumber
	&=& \frac{A}{4G} \bigg(1+f_{q}\bigg)^{\frac{1}{q}} {}_2F_{1}\bigg[\frac{2}{q}, \frac{-1}{q(n-1)}, 1-\frac{1}{q(n-1)}, -f_{q}\bigg] \,, \\ 
	&&\label{entropy mpc2}  
\end{eqnarray}
up to some integration constant.  Here  Eq.~(\ref{cons2}) gives the relation between $f_{q}$ and $A$.
\begin{figure}[h]
\begin{center}
\includegraphics[width=0.4\textwidth]{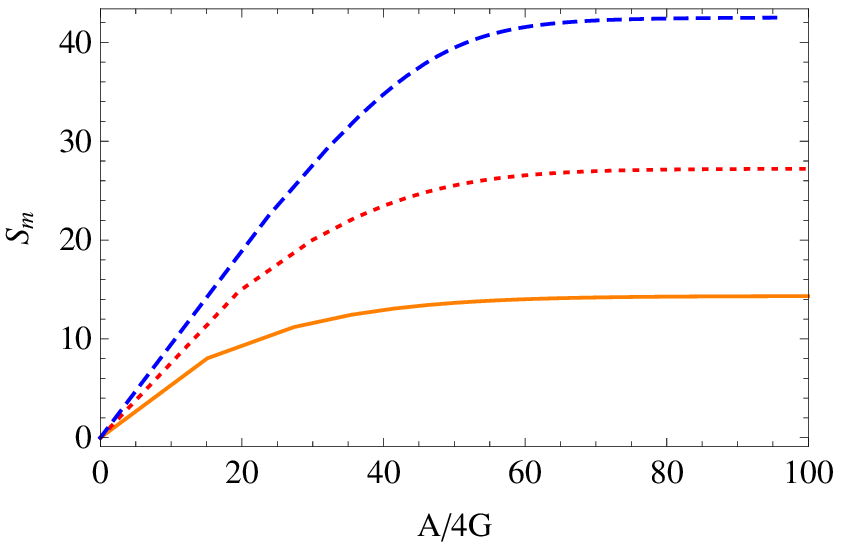}
\qquad
\includegraphics[width=0.4\textwidth]{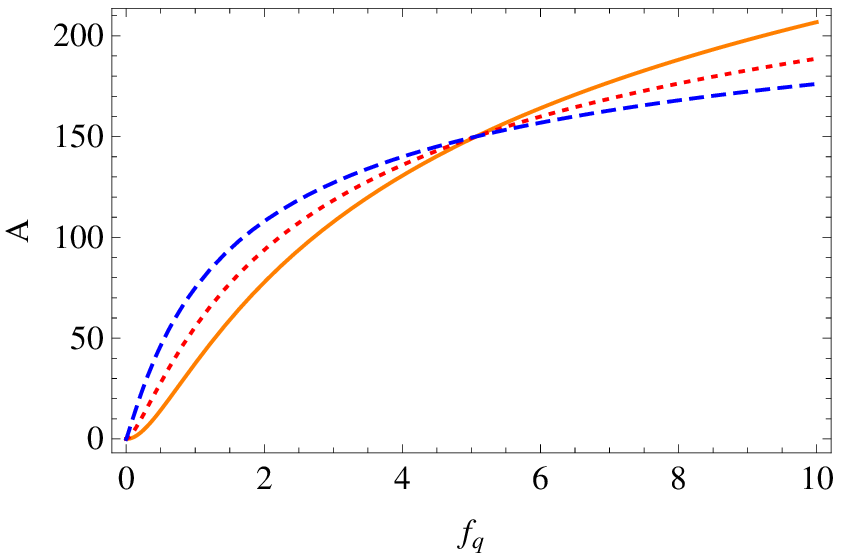}
\caption{\label{fig::empc1}Top: The entropy of the original Cardassian universe as the function of the surface area with parameter $n=0.1$ and $q = 0.5 (\text{solid}), 0.7(\text{dotted}), 1.0(\text{dashed})$. Bottom: The area of the trapping horizon with respect to function $f_{q}$ with the same parameters.}
\end{center}
\end{figure}
To illustrate the relation between  the entropy and the surface area, we plot an example with parameter $n=0.1$ in Fig.~\ref{fig::empc1}, in which it shows that when $f$ is small, the entropy satisfies the usual area law $S = A/4G$, while $f$ is large, it becomes a constant as we see in Eq.~(\ref{entropy mpc}). Also, from Eq.~(\ref{entropy mpc2}), one can obtain that 
\begin{equation}
	S_{m}\big|_{f_{q}\rightarrow \infty} = \frac{3}{8G^{2}\rho_{c}} 
	\frac{\Gamma\left(1-\frac{1}{q(n-1)}\right) \Gamma\left(\frac{2}{q} + \frac{1}{q(n-1)}\right)}{\Gamma\left(\frac{2}{q}\right)} \,,
\end{equation}
so, when $q=1$, it reduces to Eq.~(\ref{oc f}). Also, the top figure in Fig.~\ref{fig::empc1} indicates that for a given $n$, all the curves cross the point $\left(\tilde f,  \frac{3}{2G\rho_{c}}\right)$, where $\tilde f >0$ satisfies $\tilde f^{\frac{1}{n-1}}(1+\tilde f) = 1$.

\subsection{EC model} 
In this model, the function $g(\rho_{m})$ is given by
\begin{equation}\label{ec model}
	g(\rho_{m}) = \rho_{m}\exp{\left[ \left(\frac{\rho_{m}}{\rho_{c}}\right)^{-n} \right]} =  \rho_{m} e^{f_{e}(\rho_{m})}\,,
\end{equation}
where $f_{e}(\rho_{m}) = (\rho_{m}/\rho_{c})^{-n}$. Again, by using Eq.~(\ref{heatf}), the heat flow of pure matter  is given by 
\begin{equation}
	\delta Q_{m} =  \frac{T}{4G e^{f_{e}}(1-nf_{e})} \langle dA, z\rangle = T\langle dS_{m}, z \rangle\,,
\end{equation}
where the entropy is obtained by 
\begin{equation}\label{entropy ec}
	dS_{m} = \frac{dA}{4G e^{f_{e}}(1-nf_{e})} \,,
\end{equation}
which also reduces to the usual relation $dS = dA/ 4G$ in the limit of  $f_{q}\rightarrow 0$, and the entropy becomes a constant that independent of the surface area when $f_{q}$ is large enough. From  Friedmann equation (\ref{modFried}),  we also have the following constraint equation
\begin{equation}\label{cons3}
	e^{f_{e}} f_{e}^{-\frac{1}{n}} = \frac{3}{2GA\rho_{c}} \,.
\end{equation}
Thus, the area has a maximum value 
\begin{equation}
	A_{max} = \frac{3}{2G\rho_{c}}  (ne)^{-\frac{1}{n}} \,,
\end{equation}
when $f_{e} = 1/n$.  Therefore, we get the entropy by integrating Eq.~(\ref{entropy ec})
\begin{eqnarray}
\nonumber
S_{m}  &=& \frac{3}{8G^{2}\rho_{c} n }\int  e^{-2f_{e}} f_{e}^{\frac{1}{n}-1} df_{e}   \\
	&=& \frac{3}{8G^{2}\rho_{c} n 2^{1/n}} \bigg[ \Gamma(1/n) - \Gamma(1/n , 2f_{e})\bigg]  \,, \label{entropy ec2}
\end{eqnarray}
up to some integration constant.  Here  Eq.~(\ref{cons3}) gives the relation between $f_{e}$ and $A$.
\begin{figure}[h]
\begin{center}
\includegraphics[width=0.4\textwidth]{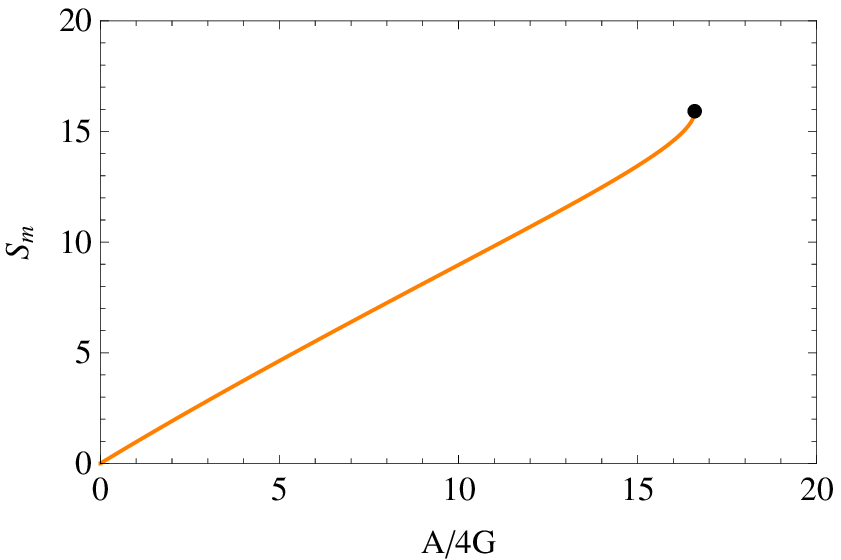}
\qquad
\includegraphics[width=0.4\textwidth]{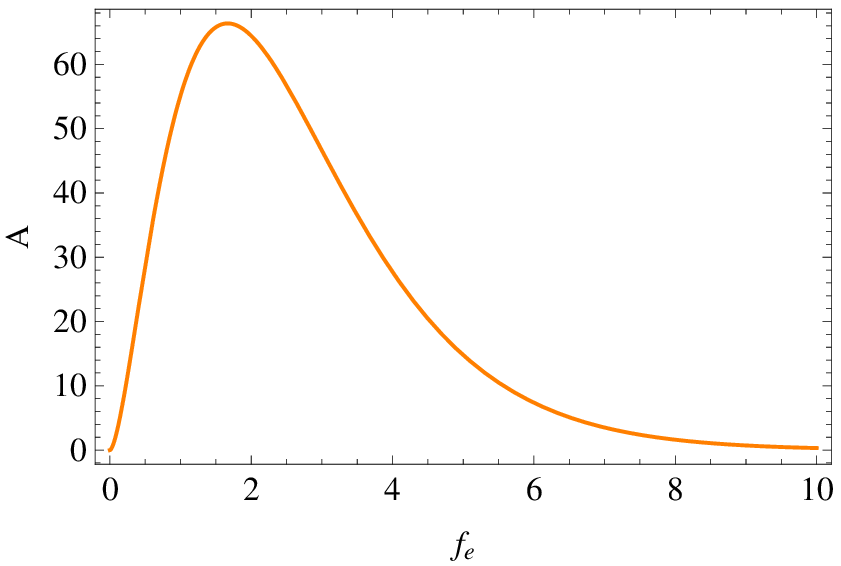}
\caption{\label{fig::eec1} Left: The entropy of the original Cardassian universe as the function of the surface area with parameter $n=0.6$ and the black point corresponds to the maximum point of the area.}
\end{center}
\end{figure}
To illustrate the relation between  the entropy and the surface area, we plot an example with parameter $n=0.6$ in Fig.~\ref{fig::eec1}, in which it shows that when $f$ is small, the entropy satisfies the usual area law $S = A/4G$, and it reaches its maximum value at $A = A_{max}$ when $f_{e} = 1/n$.  Actually, from Eq.~(\ref{entropy ec2}), one can obtain that 
\begin{equation}
	S_{m}\big|_{f_{q}\rightarrow 0} \approx \frac{3}{8G^{2}\rho_{c}} f_{e}^{\frac{1}{n}}  \approx \frac{A}{4G} \,,
\end{equation}
where we have used the relation (\ref{cons3}).

\section{Discussion and Conculsion}
In this paper, we have studied the thermodynamics of the Cardassian universe and calculated its corresponding entropy, in particular, for the OC, MPC and EC Cardassian models and the thermodynamic law is regarded as the origin of Cardassian model. We find that the entropy obeys ordinary area law on the trapping horizon when the area is small, and it becomes a constant when area is going to be large in the OC and MPC model, while it has a maximum value in the EC model. 

As we known that, the holographic principle states that all the information in the bulk should be encoded in the boundary of the bulk, so it  seems that the Cardassian universe could only contain finite information. This may lead to a question that does the information in our universe will be infinite or not? Of course, if the  ordinary area law $S=A/(4G)$ is always valid, the entropy will blow up when $A$ goes to infinite value, in contrast with the case of that in the Cardassian models,  $S$ finally becomes a constant or gets its maximum value, when $A$ is large. We will make a further study on this interesting topic \cite{fur}.

It should be noticed that, the cosmological constant will not change the area law  $S=A/(4G)$, because only the derivative $g'$ emergences in Eq.~(\ref{heatf}) or (\ref{heatf2}), but in general, different dynamic dark energy models will change the law differently. So, the method we used in this paper provides a new way to distinguash different kinds of dark energy models.

\acknowledgments

This work is supported by National Science Foundation of China grant No.~11047138, National Education Foundation of China grant  No.~2009312711004 , Shanghai Natural Science Foundation, China grant No.~10ZR1422000 and  Shanghai Special Education Foundation, No.~ssd10004.

\end{document}